\documentclass[useAMS,usenatbib]{mn2e}

\usepackage{graphicx}

\def\msun {\ensuremath{M_{\odot} \;}}

\def\arcsec {\ensuremath{^{\prime\prime}}}

\def\yr {\hbox{yr$^{-1}$}}

\def\gtrsim{\mathrel{\hbox{\rlap{\hbox{\lower4pt\hbox{$\sim$}}}\hbox{$>$}}}}
\def\lesssim{\mathrel{\hbox{\rlap{\hbox{\lower4pt\hbox{$\sim$}}}\hbox{$<$}}}}

\title[GRBs and SGRs by high energy leptons showering in blazing $\gamma$ jets]
{GRBs and SGRs by high energy leptons showering in blazing
$\gamma$ jets: are SGRs sources of EeV CRs? }
\author[D. Fargion, M. Grossi]{D.
Fargion$^{1,2}$\thanks{E-mail:daniele.fargion@roma1.infn.it}
and M. Grossi$^{1}$\thanks{E-mail:marco.grossi@roma1.infn.it}\\
$^{1}$Physics Department, Universit\'a di Roma  "La Sapienza", Pl.
A. Moro 2, 00185, Rome,  Italy\\
$^{2}$INFN, Universit\'a di Roma  "La Sapienza", Pl. A. Moro 2,
00185, Rome,  Italy}
\begin{document}

\date{}

\pagerange{\pageref{firstpage}--\pageref{lastpage}} \pubyear{2005}

\maketitle

\label{firstpage}

\begin{abstract}

The  $apparently$ huge energy budget of the gamma ray burst  GRB
990123 led to the final collapse of the isotropic fireball model,
forcing even the most skeptical to consider a beamed Jet emission
correlated to a supernova (SN) explosion. Similarly the surprising
giant flare from the soft gamma repeater SGR 1806-20 that occurred
on 2004 December 27, may induce the crisis of the magnetar model.
If the $apparently$ huge  energy  associated to this flare has
been radiated isotropically, the magnetar should have consumed at
once most of (if not all) the energy stored in the magnetic field.
On the contrary we think that a thin collimated precessing jet,
blazing on-axis, may be the source of such  apparently huge bursts
with a moderate  output power. Here we discuss the possible role
of the synchrotron emission and electromagnetic showering of PeV
electron pairs from muon bundles. A  jet made of muons may play a
key role in avoiding the opacity of the SN-GRB radiation field. We
propose a similar mechanism to explain the emission of SGRs. In
this case we also examine the possibility of a primary hadronic
jet that would produce ultra relativistic $e^{\pm}$ (1 - 10 PeV)
from $\pi - \mu - e$ or neutron decay. Such electron pairs would
emit a few hundreds keV radiation from their interaction with the
galactic magnetic field, as it is  observed in the intense
$\gamma$-ray flare from SGR 1806-20. A thin precessing jet
($\Delta \Omega \simeq 10^{-9}-10^{-10} $ $sr$) from a pulsar may
naturally explain the negligible variation of the spin frequency
$\nu=1/P$ after the flare ($\Delta \nu < 10^{-5}$ Hz). A
correlation between SGR 1806-20, SGR 1900 +14, Cygnus and the
AGASA excess of EeV cosmic rays (CRs) has been found;  we suggest
that a robust EeV signal may be detected  by AUGER or Milagro in
the near  future.


\end{abstract}

\section{Introduction}

\subsection{GRBs fireball versus thin precessing jets}

The association between gamma ray bursts (GRBs) and SNe has become
more and more convincing in the last few years after the discovery
of two main events.  In 1998, the spectroscopical and
photometrical analysis of the GRB 980425 afterglow revealed for
the first time the presence of a SN related to the GRB event
(Galama et al. 1998), and more recently SN2003dh has been detected
in coincidence with the GRB $030329$ (Hjorth et al. 2003; Stanek
et al. 2003). At least other two GRB events seem to be connected
to SNe explosions (GRB 011121/SN2001ke, Garnavich et al. 2003;
GRB031203/SN2003lw; Malesani et al. 2004). At the time of its
discovery, the GRB980425/SN1998bw association faced large
resistance among the scientific community, whose attention was
focused on the fireball scenario where this SN/GRB connection was
unacceptable. Indeed the spherical fireball  has been in the last
decade the most popular model to interpret GRBs, with the basic
assumption  that the energy is radiated  {\em isotropically}.
Consequently the output power of GRBs would be billion times
higher than that of a SN. According to the fireball model, GRBs
are produced in shocks of highly relativistic outflows with
different Lorentz factors via synchrotron radiation of electron
pair. These so-called "internal shocks" are thought to produce the
highest energy gamma emission. The consequent interaction of the
stellar ejecta with the circum-stellar medium (external shocks)
would be responsible of the multi-wavelength afterglow and its
complex time structure discovered for the first time by Beppo Sax
(Costa et al. 1997).

However isotropic explosions would require extremely high input
energies, the larger and harder the  more distant the GRBs (the so
called "Amati " law, whose beahaviour is opposite to the Hubble
law); for example, to account for the GRB 990123 emission, about
two solar masses should have been $entirely$ converted in
gamma-rays (which should be doubled if MeV neutrino pairs were
taken into account; Fargion 1999). Very massive  black holes (10 -
100 $M_{\odot}$) are needed to fuel such fireball explosions. But
the inner GRB time structure, as short as a tenth of milliseconds,
would require a Schwarzschild radius smaller than a few tens of
km, corresponding to objects of only a  few solar masses. Such a
discrepancy, added to the  SN/GRB connection, made clear the
inconsistence of the fireball scenario. Later on new families of
fireball models with a wide beamed emission have been introduced
to overcome this puzzle and reduce the energy budget (such as the
Hyper-Nova, Supra-Nova, or the Collapsar models). In these
"compromise" scenarios the outflow was thought to occur in a
"fountain" cone with opening angles as large as $\sim$ 5$^{\circ}$
- 10$^{\circ}$. However we think that even this "mild" beaming is
unable to solve the whole GRB puzzle.

In fact the beamed emission proposed in such  models reduces the
required power output by three orders of magnitude, down to
$10^{50}$ erg s$^{-1}$, nonetheless this is yet $10^6$ - $10^7$
times more intense than the observed SN power. There is no reason
to expect such a huge  unbalance between  the GRB jet and the SN
luminosities, $\dot{E}_{GRB}\simeq 10^{6}\times \dot{E}_{SN} $.
Moreover a few degrees jet  does not solve the puzzle of the low
$\gamma$ ray power detected in the GRB980425/SN1998bw event, which
finds a natural understanding assuming a  narrow jet blazing the
observer off-axis (Wang \& Wheeler 1998; Fargion 1999). Moreover a
single one-shot explosion beamed in a wide jet cannot account for
the presence of  X-Ray precursors  before the onset of a few GRB
events ($\sim 6\%$).

In this context, we have been proposing for the last decade, an
alternative scenario to the fireball  where  GRBs (as well as
SGRs) are originated by highly collimated  precessing and blazing
gamma jets with an aperture angle $\theta \sim$ 0.02$^{\circ}$ -
0.006$^{\circ}$ (Fargion $\&$ Salis 1995a, 1995b, 1995c; Fargion
1999).

The much narrower solid angle of the jet, $\Delta \Omega/\Omega
\simeq 10^{-8}$ - $10^{-9}$,  can reconcile (assuming the energy
equipartition between the powers of the SN and the jet) the puzzle
of   the "low" observed isotropic SN luminosity respect to the
apparent "huge" GRBs power (an observed  $10^{-8}$ - $10^{-9}$
ratio). The angular dynamics (a spinning and multi-precessing jet)
may explain the wide range of properties observed in different GRB
events (multiple bursts, re-brightening and bumps, X-ray
precursors, dark GRBs, X-ray flashes, low luminous off-axis GRBs;
Fargion 1999; Fargion 2003).

Within this framework, the Amati-Ghirlanda  $E_{peak} - E_{iso}$
correlation (Amati et al. 2002, Ghirlanda et al. 2005) is just the
effect of a very biased selection. At increasing distances (high
$z$) and volumes, a  larger  sample of highly collimated GRBs
becomes observable; the jets more collimated towards the observer
appear as very hard and powerful GRBs, while those blazing
off-axis would be mostly obscured and below the detectors'
sensitivity. At extreme redshifts the number of GRBs should drop
off (due to a finite age of their progenitor and the absence of SN
events) or they might correspond to the mysterious short GRB
events.

However assuming a thin one-shot jetted model leads to a
contradiction with the observations and it does not provide the
ultimate solution to the issue. In fact, such a strong collimation
would imply that the rate of one-shot beamed GRBs should be
$\Gamma_{GRB}^{beamed} \sim 10^{8}$ - $10^{9} \Gamma_{GRB}^{obs}$,
where $\Gamma_{GRB}^{obs} \sim 10^{-7} - 10^{-8}$ \yr
galaxy$^{-1}$. Given that GRB are expected to be related to SN
explosions, this value would be at least $10^2 - 10^4$ times
higher than all the observed SN rate, $\Gamma_{SN} \sim 10^{-2}$
\yr galaxy$^{-1}$. This apparent inconsistence may be solved if
one assumes that the jet structure is precessing and persistent,
with a decay lifetime much longer than the typical observed GRB
timescale ($t_{GRB} \sim 10$ s, $t_{Jet} \gtrsim 10^4$ s,
$\dot{E}_{Jet} \simeq \dot{E}_{Peak}
(\frac{t}{t_{Jet}})^{-\alpha}$, where $\alpha \geq 1$, in order to
converge to the value of the total output energy). Therefore we
argue that the jet continues to exist after the SN event, although
it is fading with time. This reduces the number of GRBs needed to
account for the observed GRB rate. The activity of the jet might
be later rejuvenated by an accretion disc, possibly fed by a
nearby compact companion. The torque perturbation created by the
companion induces  the multi-precession of the GRB jet.

This might also explain the abundance (nearly two-third) of the
so-called "orphan" or "dark" GRBs, where the brightest event
associated to the SN and its consequent afterglow may be already
faded just a few weeks or months before.  The delayed jet activity
may blaze again and it may be source of rare optical and radio
re-brightening  of the GRB, while much older and weaker sources,
may be observed thousands of years later   as SGRs, or Anomalous
X-Ray pulsars\footnote{There might be a natural extension of the
jet model to nearby low energy sources such as the X-Ray Flashes
(XRF), where a jet blazing off-axis could explain their
properties.}, provided that they are nearby and on-axis. This
occurs mostly in our galaxy and in the Large Magellanic Cloud.

Let us remind that a few  GRBs and SGRs did show comparable time
structures and spectra (Fargion 1999), even if a possible
connection between GRBs and SGRs has been often underestimated or
even rejected (Woods et al. 1999). Consequently the "necessary"
evolution of the fireball scenario into a jetted model  may be
reflected on SGRs, whose  blazing  nature challenges the
interpretation given by the {\em magnetar} model (Duncan \&
Thompson 1992; Thompson \& Duncan 1995). In fact, similarly to the
GRB 990123 fireball crisis, the very exceptional giant flare from
SGR 1806-20 occurred on 2004 December 27 is suggesting that an
{\em isotropic} magnetar  has to be questioned as the explanation
of these events for several reasons that we discuss in the
following section.

\subsection{SGRs and the puzzles of the magnetar model}

Soft Gamma Repeaters are nearby X-ray sources  localised within
the Milky Way  (excluding SGR 0525-66 in the Large Magellanic
Cloud), emitting sporadic short bursts of soft gamma photons up to
hundreds of keV. The giant flare from  SGR 1806-20 on 2004
December 27 has been characterised by an energy release which
largely exceeded all previous recorded events. It should be
noticed that the  giant flare energy exceeded  the integral sum of
all  GRBs and SGRs ever recorded. This unique event, averaged over
100 years, would make the SGRs flux much more intense than that of
GRBs ($<\Phi_{SGR}> \simeq 100$ eV cm$^{-2}$ s$^{-1}$ sr$^{-1}$).
The flare, if it was blazing far from the sun in a dark night
(contrary to the real case), would have shone (while showering and
ionising the atmosphere) like a brief visible aurora in the  sky.

 At an assumed distance of 15 kpc the isotropic energy associated with the flare would be $E = 3.5-5 \times
10^{46}$ d$^2_{15}$ ergs (Hurley et al. 2005; Schwartz et al.
2005) with a peak luminosity in the first 0.125 s equal to $L =
1.8-2.7 \times 10^{47}$ erg s$^{-1}$.  Only two other giant
outbursts of energy above $10^{44}$ ergs have been observed in
about 30 years of previous SGR activity, but never at such a huge
peak power ($\dot{E} \sim 10^{47}$ erg s$^{-1}$).  While the
giant flare energy is a tiny fraction ($10^{-4}$) of an
electromagnetic galactic Supernova, its peak power exceeds the
optical SN output by nearly the inverse ratio($\simeq 10^{4}$).
Note that the energy budget could be lowered if SGR 1806-20 were
closer, as it has been recently proposed that its distance may be
within  6.4 -- 9.8 kpc (Cameron et al. 2005). Note also that a
closer distance implies an extreme relativistic propagation of the
radio ejecta ($\gtrsim 0.6 c$).

The most popular model to interpret the properties of SGRs is the
magnetar, a neutron star with a very high magnetic field (B $\sim
10^{14} - 10^{15}$ G). This model implies an isotropic thermal
energy release like a mini-fireball, and it was born in the 90's
when fireball models were ruling the understanding of GRBs.
Surprisingly, a flare of energy 3.5 -- 5 $\times 10^{46}$
d$^2_{15}$ erg radiated isotropically would need to convert at
once almost all of the magnetar energy stored in the magnetic
field. This would imply a remarkable decrease  of the magnetic
field energy density which should appear as a prompt decrease in
the pulse period times the period derivative, $P \dot{P}$. As a
consequence, there should be a correlation between the burst
activity and the variation of the spin period, that so far has not
been observed (Woods et al. 2002). Neither the recent giant flare
showed evidence for a variability in its period
(Mereghetti et al. 2005, Woods et al. 2005 ATEL n. 407). 

Moreover the high energetics involved in this intense flare has
raised the possibility that nearby extragalactic SGRs may be
connected to short GRBs. In this case if SGRs are isotropic
phenomena, as predicted by the magnetar model, one should observe
an anisotropy in the distribution of short GRBs compared to long
ones, with the first more concentrated towards local cluster of
galaxies, especially towards
 Virgo   the closest rich cluster in the local universe.
Virgo contains about 128  spiral galaxies (Binggeli, Tamman \&
Sandage 1987), assuming that the periodicity of a giant flare in a
galaxy may be 1 in 100 years, one should expect to see 128/100
$\sim 1.3$ events/yr clustered in the area of Virgo. If short GRBs
and SGRs are the same events, the BATSE instrument in its 10 years
life time, should have detected about 13 short GRBs with a spatial
distribution  correlated to the Virgo cluster. Such a remarkable
anisotropy,  expected in the magnetar scenario, does not appear in
the BATSE GRB catalogues neither from the Virgo cluster, nor from
galaxies with extremely high star formation rate such as Arp 299
and NGC 3256 (Popov \& Stern 2005). Moreover no similar giant
flares were found in the BATSE data in nearby galaxies (within 2
and 4 Mpc) with ongoing massive star formation such as M 82, M 83,
NGC 253 and NGC 4945 (Popov \& Stern 2005). According to the
authors, the galactic rate of giant flares with energy around
$10^{46}$ erg should be less than $10^{-3}$ yr$^{-1}$.

On the other hand, a very thin, precessing  $\gamma$ Jet would
reduce the high energy budget required in an isotropic energy
release, and it  may produce at any time an exceptional bright
(on-axis) event. The rarety of such an event would imply a
negligible activity even from Virgo, still consistent with the
absence of an anisotropy in the BATSE short bursts catalogue; the
giant flare visibility may be below the threshold from clusters at
larger distances (z $>$ 0.05) (Popov \& Stern 2005).

\subsection{Structure of the paper}

We suggest that precessing and persistent jets  would better
explain the observed properties of both GRBs and SGRs, even though
several issues are yet to be addressed within this model. In this
paper we aim to improve our scenario introduced and discussed in
previous works (Fargion 1998; Fargion 1999; Fargion 2003), where
we considered jets made of relativistic GeV electrons producing
the gamma radiation by ICS,  and we try to solve some of the
problems related to this model. How can a jet of GeVs electron
pair  penetrate and propagate through  the SN envelope? How can
the collimated pairs survive the dense photon background at the
baryon column depth at the SN surface? Is it possible that a
jet is already present before the SN explosion, 
but it reaches its maximum power only after the stellar collapse?

To overcome the opacity problem  due to the photon SN background,
here we present a  scenario for GRBs where the jet is
made by muons of energy $\gtrsim$ 1 - 10 PeV. 
We assume (but we will not thoroughly discuss this hypothesis in
this paper) that in GRBs PeV muons are themselves secondaries of
ultra high energy (UHE)
neutrinos possibly  originated in a deeper and inner jet made of EeV - ZeV nucleons accelerated in the GRB/SN explosion. 
We show that the PeV muons could more easily propagate through the
dense SN radiation background   and they could also better
penetrate the outer stellar layers compared to a jet made of
electrons (or baryons). When muons decay into $\sim$ PeV
electrons, we compare the energy loss mechanisms that can produce
the observed gamma emission: synchrotron interactions due to the
stellar or galactic magnetic field,  versus ICS of $e^{\pm}$ onto
the stellar radiation background.

We discuss  a similar mechanism to explain the gamma signal from
SGRs, always assuming a primary jet made of muons. We find that
synchrotron losses (due to  the galactic  magnetic field B
$\simeq\ 2.5 \times 10^{-6}$ G) of electron pairs at $E_e \sim$
PeVs,  from the decay  of PeV muon bundles, may be the best
mechanism to originate the $\gamma$ radiation from SGRs. We
consider also that PeV muons may be secondaries of EeV nucleons
originated via photopion interactions, since we find a possible
link between the anisotropy in the EeV CR distribution near the
galactic center observed by AGASA (Hayashida et al. 1999) and
SUGAR (Bellido et al. 2001), and the position of SGR 1806-20. If
this correlation is not fortuitous this maybe the first evidence
of a connection between EeV CRs and the SGR activity. The
possibility of detecting CRs (nucleons and TeV photons) and
neutrinos from this giant flare in AUGER, MILAGRO and AMANDA is
also discussed.

 The paper is
organized as follows: in \S  2 we briefly summarise the main
characteristics  of the
precessing jet model, 
in \S 3 we discuss the problem of the opacity of electrons
propagating through a dense supernova gas photon, and we introduce
a primary muon jet model, in \S 4 we discuss the possible
mechanisms that can lead to the gamma emission from secondary
electron pair Jets, in \S 5 we apply the same model to SGRs, and
we discuss the correlation between EeV CRs and SGR 1806-20, and
finally in \S 6 we present our final discussion and conclusions.


\begin{figure} \begin{center}
\includegraphics[width=8cm]{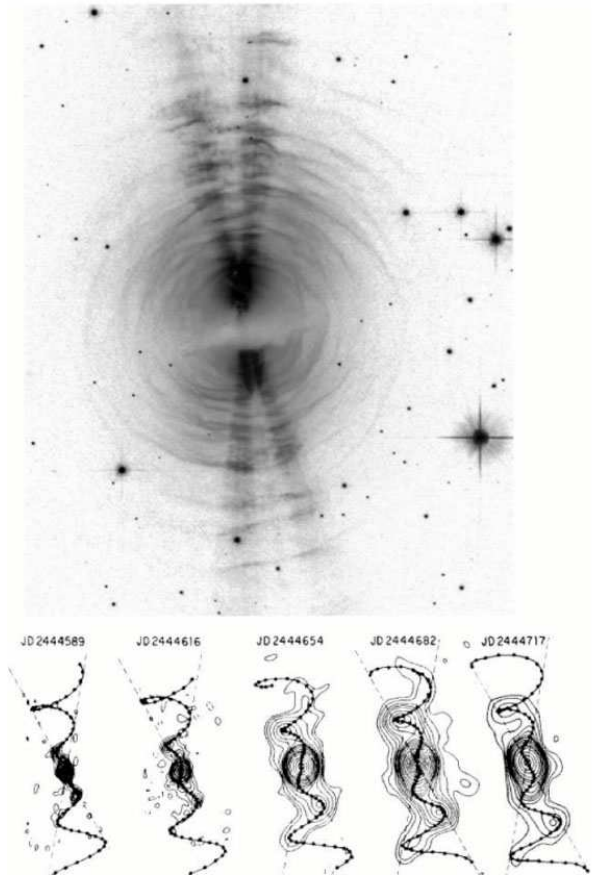}
\end{center}
\caption{{\em Up:} The Egg Nebula, whose shape might be explained
as the conical section of a twin precessing jet interacting with
the surrounding  cloud of ejected gas. {\em Down:} The observed
structure of the outflows from the microquasar SS433. A kinematic
model of the time evolution of two oppositely directed precessing
jets is overlaid on the radio contours (from Blundell \& Bowler
2005).} \label{SS433}
\end{figure}

\section{The Precessing Jet  Signatures}

Precessing jets have been already introduced to explain the
properties of other astrophysical objects. Well known
micro-quasars, such as 1915-16 and SS 433, are the best candidate
to show the multi-precessing jet evolution in nearby galactic
objects (see Fig. \ref{SS433}). Other well known sources as  CH
Cyg, Cyg X-3 and GRO J1655-40 
show precessing jets whose properties remind, at a smaller scale,
those of larger active galactic nuclei such as M87 or Cygnus A.
Bipolar planetary nebulae as the Sa2-237 (Schwarz, Corradi \&
Montez 2002) hidden in molecular clouds or the Egg Nebulae (see
Fig. \ref{SS433}), also show  features that give evidence of a
twin double cone possibly made by a precessing jet. At a lower
luminosity, jets are used to explain the properties of two
accreting white dwarfs (Zamanov et al. 2005) exhibiting collimated
outflows with a morphology well fit by a precessing jet model.
Outflows collimated in jets are also observed  in Herbig Haro
protostellar objects (HH 40, HH 34), and in the  large protostar
IRAS 16547-4247.

According to Mirabel (2004) there is an universal mechanism for
the production of relativistic jets in accreting black holes that
would unify AGN, GRBs, microquasars and Ultra Luminous X-ray (ULX)
sources.

In our model we assume that GRBs and SGRs originate from the same
process and they represent  the early and the late stages of the
evolution of a precessing jet. The jet is possibly 
fuelled by the SN event for a GRB and by an 
accretion disc or a companion in the case of SGRs. 
In SGRs the close encounter between a compact source with a fading
jet and the companion star, may strip matter from the latter,
that, falling into the accretion disc, leads to an increase or a
revival of the jet power. The  presence of a compact star (white
dwarf or a neutron star) would be the best candidate  to bend the
jet before  and after a SN in GRBs.

Therefore we propose a model of a jet that is  not static, but we
assume that it is spinning as the dense central object that hosts
it (pre-SN star or NS), and precessing, even at random, due to the
presence of a binary companion (or an asymmetric accretion disc ).
A nutation mode is superposed to the precession circle. The
spinning motion provides  a short duration  of the GRB (or SGR)
event, while precessing modes  move the jet in and out of the line
of sight leading to a re-brightening or a quasi periodic activity.

The temporal evolution of the angle between the jet direction and
the rotational axis of the object, $\theta_1(t)$, can be expressed
as

\begin{equation}
\theta_1 (t) = \sqrt{\theta_x^2 +  \theta_y^2}
\label{GRB_Equation}
\end{equation}

with \\

$ \theta_x(t) =  \sin (\omega_b t + \phi_{b}) + \theta_{psr} \cdot
\sin( \omega_{psr} t + \phi_{psr}) + \theta_s \cdot \sin (\omega_s
t+ \phi_{s}) + \theta_N \cdot \sin (\omega_N t +
\phi_N) + \theta_x(0)$ \\

 $\theta_y(t) =  \theta_a \cdot \sin \omega_0 t  + \cos ( \omega_b t + \phi_{b})+ \theta_{psr} \cdot \cos (\omega_{psr} t +
\phi_{psr}) + \theta_s \cdot \cos (\omega_s t+ \phi_{s}) +
\theta_N \cdot \cos (\omega_N t + \phi_N)) + \theta_y(0)$ \\

$\gamma$ is the Lorentz factor of the relativistic particles (half
PeV electrons).
$\theta_{psr}$, $\theta_b$, $\theta_N$ are respectively the
maximal opening angles due to the spinning of the star, the
perturbation due to the companion, and the nutation motion of the
multi-precessing jet axis. The arbitrary phases $\phi_b$,
$\phi_{psr}$, $\phi_N$, for the binary, spinning pulsar and
nutation, are able to fit the complicated features of the GRB
light curves in most of the observed events. The additional
parameters have been introduced to model the light curve of the
SGR giant flare. $\theta_a$ is the bending angle of the entire
cone precessing with a frequency $\omega_a$, corresponding to a
period of nearly a dozen years (related to "intermittent" SGRs
activity)\footnote{The real binary angular velocity might be very
slow: in this case $\omega_a$ might play the role of the the
binary angular velocity while the other ones ($\omega_{b}$,
$\omega_{N}$) are shorter precessing and nutating frequencies due
to the asymmetry of the spinning jet.}; $\theta_s$ is a possible
inner pulsation of the pulsar at frequency $\omega_s = 25$ rad/s;
$\theta_N$ and $\omega_N $ refers to the nutation motion of the
jet, and $\theta_x(0)$, $\theta_y(0)$
are fine tuned phases able to reasonably  reproduce the observed
light curve of the giant flare.

\begin{figure}
\begin{center}
\includegraphics[width=6.5cm,height=10.5cm]{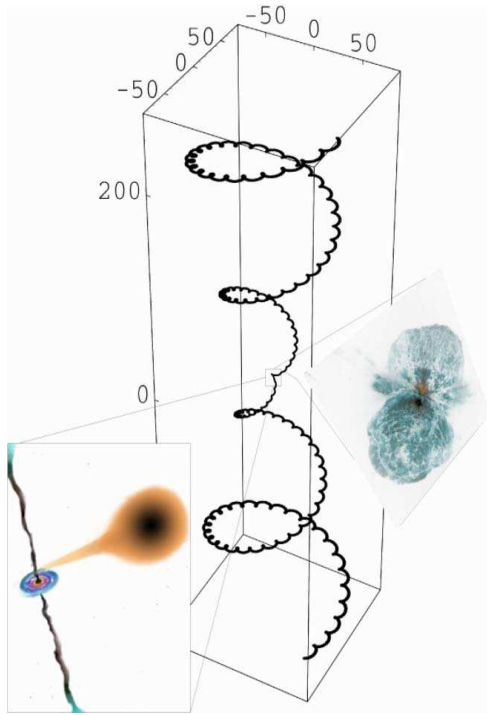}
\end{center}
\caption{A possible 3D structure of the precessing jet  obtained
with the parameters displayed in Table 1; at its centre the
"explosive" SN-like event that here we have represented with the
Eta-Carinae lobes, or a steady binary system system where an
accretion disc around a compact object powers a colimated
precessing jet.} \label{spinning_GRB}
\end{figure}

In Fig \ref{spinning_GRB} we show the results of our models for
the parameters displayed in Table \ref{jet_parameters}.

\begin{figure}
\begin{center}
\includegraphics[width=6cm,height=6cm]{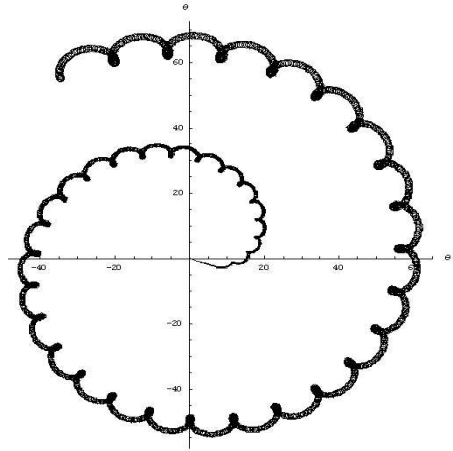}
\end{center}
\caption{The jet's 2D projection along a plane orthogonal to the
vertical axis. Such a configuration of the jet has been obtained
with the parameters displayed in Table 1.} \label{spinning_GRB_2D}
\end{figure}

\begin{table}
\begin{tabular}{lll}
\hline \hline
  $\gamma = 10^9$  & $\theta_a=0.2$ & $\omega_a =1.6 \cdot 10^{-8}$ rad/s\\
  $\theta_b=1$ &  $\theta_{psr}$=1.5 $\cdot 10^7$/$\gamma$ & $\theta_N$=$5 \cdot 10^7$/$\gamma$ \\
$\omega_b$=4.9 $\cdot 10^{-4}$ rad/s &  $\omega_{psr}$=0.83 rad/s
& $\omega_N $=1.38 $\cdot 10^{-2}$ rad/s \\
$\phi_{b}=2\pi - 0.44$ &$\phi_{psr}$=$\pi + \pi/4$ & $\phi_N$=3.5
$\pi/2 + \pi/3$ \\
$\phi_s \sim \phi_{psr}$ & $\theta_s$=1.5 $\cdot 10^6$/$\gamma$ & $\omega_s = 25$ rad/s \\
 \hline \hline
\end{tabular}
\caption{The parameters adopted for the jet model represented in
Fig. \ref{spinning_GRB}} \label{jet_parameters}
\end{table}

We have shown already (Fargion 1999, Fargion 2003) that by varying
the parameters of this simple model one can fit the profiles of
the GRB events. Moreover we proved (Fargion $\&$ Salis 1995a,
1995b, 1995c, Fargion \& Salis 1998, Fargion 1999) that the GRB
$\gamma$ observed spectrum may be well fit by the Inverse Compton
Scattering of  relativistic electrons with a
 power-law or a monochromatic spectrum.


\section{The nature of the GRB jets}

\subsection{The opacity  of the SN radiation to an
electron pair jet}

One important issue is whether the jet is born during the SN event
or if it is already present.

If the jet is created during or immediately before the SN
explosion as soon as the GRB emission is detected, it is unlikely
that a jet powered for the typical duration of a long burst $\sim$
20 s could penetrate a giant star. It would rather dissipate its
energy in the stellar envelope. Moreover it would take about 1000
s for the jet  to penetrate a red supergiant and break the stellar
surface (Rosswog 2003).

Recent theoretical models assume that  the progenitor of a GRB is
a massive  star which has already lost its H envelope before core
collapse, given the correlation between  GRBs and supernova of
type Ic (i.e. core-collapse but no hydrogen and helium lines).
These type of stars are usually identified as Wolf-Rayet and are
characterised by initial masses of  about 25 - 40 \msun (Maeder \&
Conti 1994). The heavy mass loss rate that distinguishes these
objects reduces the mass of the star by losing the outer hydrogen
and helium layers, leaving a carbon/oxygen core (Matheson 2004).
In general they are expected to have less massive envelopes than
SNe Ib. Typical masses are around 16 to 18 \msun but the range is
 wide: from 5 \msun to 48 \msun. Therefore W-R stars are the low
mass descendants of previously massive O stars (Maeder 1990).

Radii are very difficult to determine for these stars, especially
because of the strong mass loss that makes it difficult to define
what is  the stellar 'surface'. However,  few estimates exist for
eclipsing binaries, where the radius is found to be  11
$R_{\odot}$ for the late type CQ Cephei (HD 214419, WN7+O9) and
$\sim 3 R_{\odot}$ for the earlier V444 Cygni (HD 193576, WN5+O6).
There appears to be a  correlation between large radii and
late-type WR stars (where the Hydrogen lines are detected), while
 early type WR stars (no Hydrogen lines) show smaller radii.

 Therefore such a smaller progenitor with a thinner
stellar envelope would have more favorable conditions for the jet
to emerge from  the stellar surface (as for the  GRB 021004, see
Starling et al. 2005). GRB 031203, GRB 030329, and the most
popular GRB event related to SN 1998bw, all show spectra typical
of a type Ic SN.

Another possibility would be that the jet is already  present
before the SN event, as one can see in young stellar sources,
whose jet are well observable and often in precession,  such as
Herbig Hero objects.

Although there is no generally accepted mechanism able to explain
the making of a thin jet structure, there seems to be a link
between jets and accretion discs.

\subsection{Muon jets to overcome the SN-GRB opacity}

\begin{figure}
\begin{center}
\includegraphics[width=8cm,height=8cm]{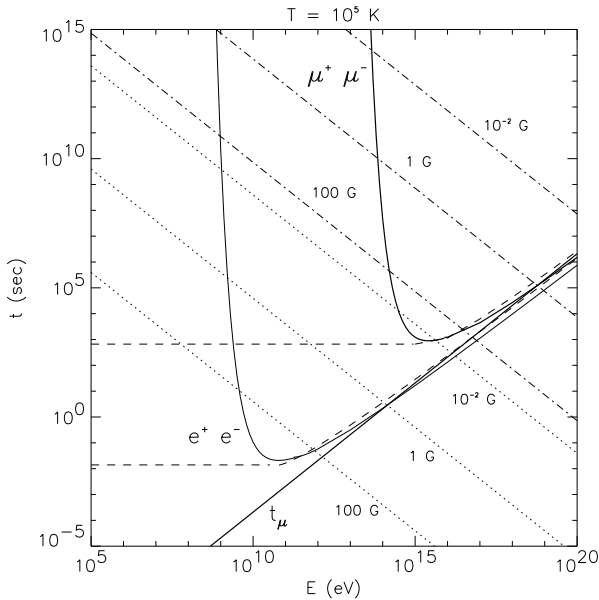}
\caption{The electron and muon interaction lengths. The {\em
dashed-dotted} and {\em dotted} lines correspond to the
synchrotron energy loss distance (for muons  and electrons
respectively) for different values of the magnetic field: 100 G, 1
G and $10^{-2}$ G. The {\em straight solid} line labelled
$t_{\mu}$ indicates the muon lifetime. The {\em dashed} lines
indicate the IC interaction lengths for muons and electrons.
Finally the two {\em solid} curves labelled $\mu^+ \mu^-$ and $e^+
e^-$ correspond to the attenuation length of high energy photons
producing lepton pairs (either $\mu^{\pm}$ or $e^{\pm}$) through
the interaction with the SN radiation field.  We have assumed that
the thermal photons emitted by the star in a pre-SN phase have a
black body distribution with a temperature $T \simeq 10^5$ K.
Assuming a radius $R \sim 10 \, R_{\odot}$, we are considering a
luminosity of $L_{SN} \simeq  2.5 \cdot 10^{41}$ erg  s$^{-1}$.
Around $10^{15} - 10^{16}$ eV muons decay before losing energy via
IC scattering with the stellar background or via synchrotron
radiation.} \label{interaction_length}
\end{center}
\end{figure}

In our previous models of GRB emission  we  assumed a jet made of
electron pairs of energy around 1 - 10 GeV  producing the
gamma-ray emission by IC onto thermal photons (Fargion 1995a,
1995b, 1995c; Fargion 1998; Fargion 1999; Fargion 2003). However,
the main difficulty for a jet of GeV electrons is that their
propagation through the SN radiation field is highly suppressed.

Indeed high energy electrons and positrons ($E_e > 1$ GeV) may
inverse Compton scatter the UV/optical  photons emitted by the
star to high energies leading to an electron-photon cascade. The
interaction length of $e^{\pm}$ of energy $E_e$, in an isotropic
photon gas with density $n(\epsilon)$, is given by (Protheroe
1986)

\begin{equation}
\lambda^{-1}_{IC} = \frac{1}{2} \int_0^{\infty} n(\epsilon)
\int_{-1}^{1} \sigma_{e \gamma} (\omega) (1 - \beta \cos \theta) d
(\cos \theta) d\epsilon \label{eIC}
\end{equation}

where $\theta$ is the angle between the electron and photon
directions in the laboratory frame and $\omega = (\epsilon E/m_e
c^2) (1 - \beta \cos \theta)$ is the photon energy in the electron
rest frame.

In Fig. \ref{interaction_length} we show the IC interaction length
for electrons and muons ({\em dashed lines}) scattering  a
black-body radiation field at a temperature T $= 10^5$ K. In the
Klein-Nishina limit we have used the asymptotic formula given by
Gould \& Raphaeli (1978). From Fig. \ref{interaction_length} one
can see that electrons at $E_e \gtrsim 10$ GeV can not propagate
for more than 10$^{-3}$ s due to the scattering with the
optical/UV emission. Moreover at $10^{-3}$ s ($\sim$ 300 km) from
the stellar surface the baryon density would also be able to
reduce and block the UHE electron pair propagation. This
represents a strong constraint for a jet model made of $e^{\pm}$.

UHE muons ($E_{\mu} \gtrsim$ PeV) instead are characterised by a
longer interaction length either with the circum-stellar matter
and the radiation field, since the $\mu - \gamma$ cross section
scales as the inverse square of the mass of the lepton involved
(in the Thomson regime). In the Klein-Nishina regime at E $>
10^{17}$ eV the interaction lengths for muons and electron becomes
comparable (see Fig. \ref{interaction_length}).

For this reason here we discuss the possibility  that the primary
particles of the jet may be muons with $E_{\mu} \sim 10^{15}$ --
$10^{16}$ eV, which can more easily escape from the stellar
interior. A jet of primary electrons at the same energy would have
a similar IC interaction length but their escape from the stellar
surface would be obstructed by both the baryon load (i.e.
electromagnetic interactions inside the matter) and by the
synchrotron opacity which would be dominant at those energies for
a magnetic field between 1 and 100 Gauss (see {\em dotted lines}
in Fig. \ref{interaction_length}).  From this starting point we
investigate the mechanism able to produce the observed gamma
radiation in GRBs from muon jets.

We assume that the star where the GRB occurs is a WR-like object
with a characteristic radius $R_{\star} \sim 10 R_{\odot}$. When
the muon jet breaks through the surface of the star it may
interact either with the magnetic field of the star or inverse
Compton (IC) scatter its optical -- UV photon field.

The importance of the synchrotron process depends on the magnetic
field configuration around the jet and the star. In the case of a
WR star, the magnetic field on the surface may be as high as 100 G
 and assuming a $\frac{1}{r^3}$ dependence, one can estimate the
field intensity at a distance $r$.

As an example, for an initial muon at $E_{\mu} \sim 10^{16}$ eV
and a magnetic field of roughly 1 Gauss, synchrotron photons would
be emitted  at

\begin{equation}
E_{\gamma}^{sync} \sim 4  \times 10^5 \left(\frac{E_{\mu}}{
10^{16} \: eV} \right)^2 \left(\frac{B}{1\; G} \right) \: eV
\end{equation}

and the characteristic interaction length is

\begin{equation}
 t^{sync}  = \frac{E}{\frac{4}{3}
\sigma_T^{\mu} c \gamma_{\mu}^2 U_B}  \simeq  5 \times 10^7
  \left(\frac{E_{\mu}}{10^{16} eV} \right)^{-1}
\left(\frac{B}{1\; G} \right)^{-2} \: s
\end{equation}


It follows  (see Fig. \ref{interaction_length}) that for $B < 100$
G the synchrotron interaction distance scale ({\em dashed-dotted
lines}) is much longer than the IC ({\em upper dashed line}) and
the muon decay length ({\em straight solid line}), therefore muons
can propagate outside the star without significant synchrotron
losses. Only for a magnetic field around $10^3$ G, muons would
lose most of their energy to synchrotron emission of photons with
$ E_{\gamma} \sim 400$ MeV, but such high magnetic fields in stars
are yet to be confirmed.

Nevertheless for $E_{\mu} \sim 10^{16}$ eV even the interaction
length of the IC scattering onto the stellar radiation field is
larger than the muon lifetime (Fig. \ref{interaction_length}),
therefore such muons decay into electrons (in roughly 100 s)
before their energy is dissipated via IC or synchrotron. As a
consequence,
 a jet of muons would be able to escape from the
star and propagate for about 100 light-seconds before decaying
into electrons. To understand the fate of such high $e^{\pm}$, one
needs to compare again the IC and synchrotron interactions to
establish which is the mechanism that most affects their loss of
energy as they propagate outside the stellar surface.

\section{GRBs by electrons showering in $\gamma$ jets}

The main source of the thin gamma jet is the ultra high energy
electron pairs showering into thin $\gamma$ jets. PeV electrons
and positrons from muon decay may either inverse Compton scatter
the UV/optical photons emitted by the star, leading to an
electron-photon cascade, or lose energy emitting synchrotron
radiation in galactic magnetic fields.

From Fig. \ref{interaction_length} it appears that synchrotron
interactions  are less favoured compared to ICS because their
distance scale is larger. In fact, even assuming that intensity of
the magnetic field of the pre-SN star, $B$, decreases as r$^{-2}$
(rather than r$^{-3}$),  at
 100 light seconds from the star, $B$ is low enough ($< 10^{-2}$ G, if the field on the surface of the star is $\sim 100$ G) that
 electrons survive synchrotron losses,  and IC constitutes
 the dominant energy loss mechanism (see in Fig. \ref{interaction_length}
the synchrotron energy loss distance for $B = 10^{-2}$ G and the
IC interaction length for electrons at energies around $10^{15}$
eV).

Compton interactions for PeV electrons occur in the Klein-Nishina
regime, since the ambient photon energy in the rest frame of the
electron is $\omega \gg m_e c^2$, and the $e^{\pm}$ pair tend to
transfer most of their energy (about $\frac{1}{2} E_e$) to the
background photons. The propagation of the hard $\gamma$-rays
created is attenuated by the collisions with the stellar radiation
background that leads to pair production ($\gamma + \gamma
\rightarrow e^+ e^-$). Compton scattering and pair production
interactions concur to produce a complicated "electromagnetic
cascade" which appears to be the favoured mechanism for the PeV
electrons to lose most of their energy (see also Fargion \&
Colaiuda 2004).

The attenuation length for a hard photon of energy E, traversing a
black-body photon gas of energy $\epsilon$ and density
$n(\epsilon)$ is given by a relation similar to Eq. \ref{eIC}

\begin{equation}
\lambda^{-1}_{pair} = \frac{1}{2} \int_0^{\infty} n(\epsilon)
\int_{-1}^{1} \sigma_{\gamma \gamma} (s) (1 - \beta \cos \theta) d
(\cos \theta) d\epsilon \label{egamma}
\end{equation}

where $\sigma(s)$ is the  pair production cross section for
photons with center of mass energy $\sqrt{s}$, and $s = 2 \epsilon
E (1 - \beta \cos \theta)$. $\lambda_{pair}$ has been derived
following Brown et al. (1973) and it is shown in Fig.
\ref{interaction_length} ({\em solid} curve labelled $e^+e^-$) as
well as the interaction length of the process $\gamma \gamma
\rightarrow \mu^+ \mu^-$.

\begin{figure}
\begin{center}
\includegraphics[width=8.5cm,height=8.5cm]{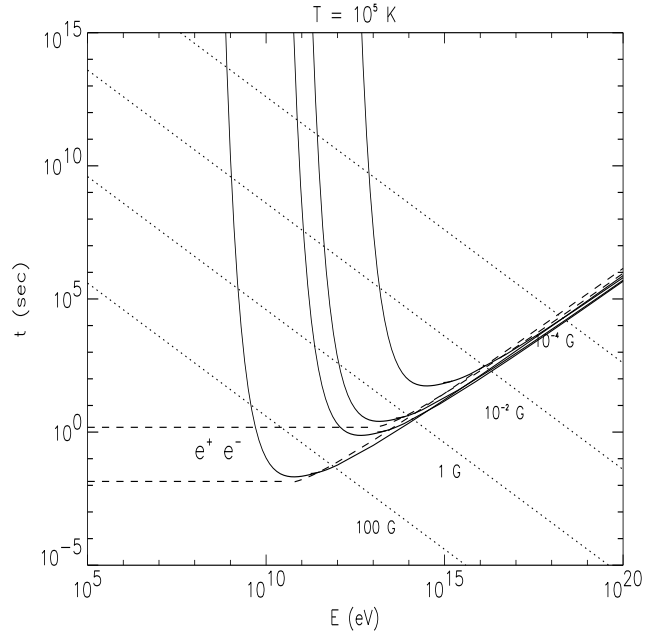}
\caption{The evolution of the $e \gamma$ ({\em dashed line}),
$\gamma \gamma$ ({\em solid line}) interaction length as a
function of time. Here we have plotted the curve at $t$ = 0 s, 100
s (when $\mu^{\pm}$ decay into $e^{\pm}$), 200 s (the lapse time
of ten $e \gamma$ bounces), and 1000 s (from left to right). The
IC and pair production interaction lengths are compared to the
synchrotron one for B = 100, 1, $10^{-2}$, $10^{-4}$ Gauss.}
\label{interaction_length_time}
\end{center}
\end{figure}

To summarise, in the $e - \gamma$ collisions (ICS), PeV electrons
transfer half of their energy to the ambient photon, producing a
hard $\gamma$-ray which then creates an $e^+ e^-$ pair due to the
interactions  with the photon stellar background ($\gamma - \gamma
\rightarrow e^+ e^-$). However, as the electron propagates
outward, the photons encountered are progressively redshifted, and
the interaction length of the $e^{\pm}  \gamma$ and $\gamma
\gamma$ scattering is continuously changing with distance and time
as we show in Fig. \ref{interaction_length_time}. We are assuming
that the energy of the photons is redshifted by a factor $(1 -
\beta \cos \theta)$, where $\tan \theta = R_{\star}/(l_e +
R_{\star})$, with $l_e$ being the distance covered by the electron
at a certain time $t$, and $R_{\star}$ is the WR star radius.
The multiple photon-electron cascade continues during the electron
propagation.  For an initial PeV electron, we consider that about
10 collisions are needed for its energy to be degraded to $E_e
\sim$ 500 GeV -- 1 TeV (see also Fargion \& Colaiuda 2004). The
lapse time for ten bounces is approximately $10 \times 10$ s
$\simeq 100$ s. At this energy, the photon produced at $E_{\gamma}
= \frac{1}{2} E_e$ is able to "escape" without interacting anymore
with the stellar background (Fig. \ref{interaction_length_time}).
This mechanism would imply a hard $\gamma$  emission (TeV,
sub-TeV) from GRBs.

 Seven GRBs with emission beyond 100 MeV have been
detected by EGRET on board of the Compton Gamma Ray Observer
(CGRO) (Schneid et al. 1992), including the long-duration burst
($\sim 5000$ s) GRB 940217 whose peak emission has been measured
at $\sim$ 18 GeV (Hurley et al. 1994). In principle one cannot
exclude that GRBs may emit at TeV energies, even though the
propagation of these photons would be suppressed by the presence
of the infrared (IR) extragalactic background, ($\gamma_{TeV} +
\gamma_{IR} \rightarrow e^+ e^-$), so that only photons from
relatively nearby sources ($< 100 - 300$ Mpc) could be detected. A
few claims of GRB detections in the TeV and sub TeV range have
appeared in the literature in the last few years. Such high energy
signals have been claimed to be observed by
Milagrito   in one out of 54 GRBs from the BATSE catalogue (Atkins et al. 2000). 
Another claim of possible sub-TeV emission  from GRB 971110 has
come from the GRAND project (Fragile et al. 2004). However one has
to keep in mind that all these results are not considered as firm
detections. It follows that even if we may expect a very high
energy ($\sim$ TeV) "precursor", it would be very difficult to
observe it
because of the IR cut-off.

Going back to the keV - MeV emission, in this scenario it would
result from  the continuous IC scattering of the same electrons,
when they have
reached lower energies. 
In fact, after they have been degraded to few hundreds GeV, the
electrons keep interacting with the stellar background but the
process is not anymore in the Klein-Nishina regime. These
collisions lead to the creation of photons of energy  given by

\begin{equation}
E^{IC}_{\gamma} = \frac{4}{3} \gamma^2 \epsilon (1 - \beta \cos
\theta)
\end{equation}

where $\theta$ is the scattering angle between the photon and the
electron\footnote{A similar showering process has been discussed
for UHECR from BL Lac sources (Fargion \& Colaiuda 2004)}. For $tg
\theta =  R_{\star}/(l_e + R_{\star})  \ll 1$, $(1 - \beta \cos
\theta) \sim \theta^2$ and the equation reduces to

\begin{equation}
E^{IC}_{\gamma} \simeq 4.5 \cdot 10^{7} \left( \frac{E_e}{100 GeV}
\right)^2 \left( \frac{\epsilon}{8.6 eV} \right) \left(
\frac{\theta}{10^{-2} rad} \right)^2 eV
\label{IC_rad}
\end{equation}

where the $\theta^2$ factor accounts for the apparent redshift in
the electron frame. The energy loss rate in this case has the form

\begin{equation}
\frac{dE}{dt} = - \left( \frac{4}{3} \frac{\sigma_T a T^4}{m_e^2
c^4} \right) E^2 = - b E^2
\end{equation}

 and the energy decrease as a function of time is given by

\begin{equation}
E(t) = (E(t_0)^{-1} + b t)^{-1}.
\end{equation}

Thus the collisions with the ambient photons rapidly lower the
energy of the electrons. When their energy is reduced to $E_e = 10
GeV$, from Eq. \ref{IC_rad} one obtains approximate photon
energies of 450 KeV $\left( \frac{E_e}{10 \, GeV}  \right)^2
\left( \frac{\epsilon}{8.6 eV} \right) \left(
\frac{\theta}{10^{-2} rad} \right)^2 $.


\section{The SGR 1806-20 Giant flare by a blazing gamma jet}

An isotropic release of energy, as predicted by the magnetar
model,  shows some inconsistencies when applied to the properties
of the giant flare from SGR 1806-20, as we have pointed out in \S
1.3. There are two more issues related to the magnetar
interpretation that we want to discuss in this section.

First, how can a spherically symmetric explosion justify the
peculiar features observed in its light curve? A precursor burst
occurred 143 s before the initial onset, lasting slightly longer
than 1 s, and it was followed by a very intense spike with a very
short timescale (0.7 sec), the blazing giant flare that saturated
most GRB detectors. After t = 61 until 170 ms the signal "{\em
decreased gradually with oscillatory modulation, which suggests
repeated energy injections at $\sim$ 60 ms intervals}" (Terasawa
et al. 2005). Soon after the peak of the emission, between 400 and
500 ms, several humps were detected in the decay profile from both
Swift (Palmer et al. 2005) and GEOTAIL (Terasawa et al. 2005).
Then the profile declines again until $t$ $\sim$ 300 s, with the
light curve showing signs of pulsations with a 7.56 s period. An
increase in the emission seem to occur  again around t $\sim$ 400
s, with a long bump peaking at $t$ $\sim$ 600 - 800 s (Mereghetti
et al. 2005). An even longer tail seems to appear from the data,
which lasted until 1000 - 2000 sec after the first trigger
(Mereghetti et al. 2005).


Secondly, does the relation between $P \dot{P}$ and $B^2$,
predicted by  the magnetar model, find a confirmation from the
observations? The magnetar scenario assumes  a neutron star with a
huge magnetic field that is estimated by observing the rotational
energy loss (mainly  due to the dipole moment radiation) and the
rate of rotation variability.

Let us calculate these quantities for SGR 1806-20 assuming a
simple dipole model. The rotational energy and its derivative (for
a nominal period of $P_{SGR}= 7.56 s$) are:

\begin{equation}
E_{rot}=  \frac{1}{2} I_{NS} \omega^2 \simeq 3.6 \cdot 10^{44}
P^{-2} \left( \frac{I_{NS}}{10^{45} g \, cm^2} \right) erg
\label{Erot}
\end{equation}

\begin{equation}
\dot{E}_{rot} = I_{NS} \omega \dot{\omega} =  4 \cdot 10^{46}
\dot{P} P^{-3}  \left( \frac{I_{NS}}{10^{45} g \, cm^{2}} \right)
\, erg \, s^{-1}
\end{equation}

However from the dipole radiation formula one finds:

\begin{equation}
\dot{E_{rot}}= -\frac{(2 \pi)^4 B_{NS}^2 R_{NS}^6 }{6 c^3 P^4}
\cdot sin^2(\theta)
\end{equation}

From the last two equations one derives that the magnetic field is
related to the period $P$ and its derivative $\dot{P}$ as follows:

\begin{equation}
B = 3.2 \cdot 10^{19} (P \dot{P})^{\frac{1}{2}} \, G
\end{equation}

Mereghetti et al. (2005) have shown the long term  $\dot{P}$
variation of SGR 1806-20 between 1993 and October 2004, using data
from ASCA, RXTE, Beppo Sax and XMM. The period derivative
increased from $8 \times 10^{-11}$ s s$^{-1}$ (before 1998;
Kouveliotou et al. 1998) to $(5.49 \pm 0.09) \times 10^{-10}$ s
s$^{-1}$ (between 1999 and 2004), while the period $P$ changed
from 7.47 s, to 7.56 s. This implies that the magnetic field has
changed from

\begin{equation}
B_{in}=  7.8 \cdot 10^{14} \left( \frac{P}{7.47 \, s}
\right)^{\frac{1}{2}} \left( \frac{\dot{P}}{8 \cdot 10^{-11} \, s
\, s^{-1}} \right)^{\frac{1}{2}} \, G
\label{Bin_versus_P}\end{equation}

to

\begin{equation}
B_{fin}=  2.06 \cdot 10^{15} \left( \frac{P}{7.56 \, s}
\right)^{\frac{1}{2}} \left( \frac{\dot{P}}{5.49 \cdot 10^{-10} \,
s \, s^{-1}} \right)^{\frac{1}{2}} \, G \label{Bfin_versus_P}
\end{equation}


it follows that $P \dot{P}$ has increased roughly of a factor 7,
as the square of the magnetic field, $B^2$. According to these
values, the corresponding  total energy  $E \simeq \frac{1}{6}B^2
R_{NS}^3$ must have changed from

\begin{equation}
E_{in}=  1.0 \cdot 10^{47} \left( \frac{P}{7.47 \, s} \right)
\left( \frac{\dot{P}}{8 \cdot 10^{-11} \, s \, s^{-1}} \right)
\left( \frac{R_{NS}}{10^6 \, cm} \right)^3  erg
\label{Uin_versus_P}\end{equation}

to

\begin{equation}
E_{fin}=  7.1 \cdot 10^{47} \left( \frac{P}{7.56 \, s} \right)
\left( \frac{\dot{P}}{5.5 \cdot 10^{-10} \, s \, s^{-1}} \right)
\left( \frac{R_{NS}}{10^6 \, cm} \right)^3  erg
\label{Ufin_versus_P}
\end{equation}

\begin{figure} \begin{center}
\includegraphics[width=6.5cm,height=6.cm]{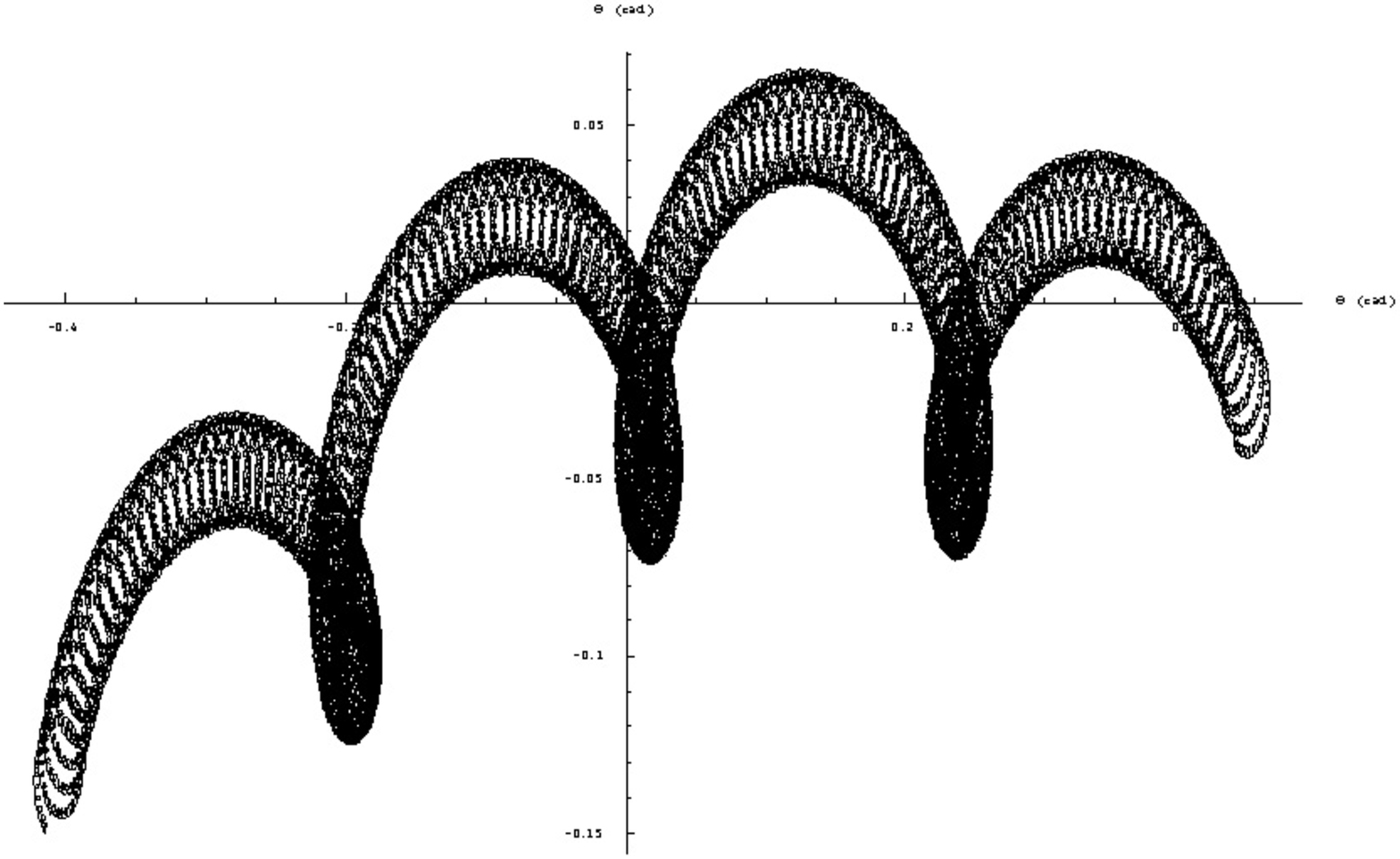}
\includegraphics[width=6.5cm,height=6.cm]{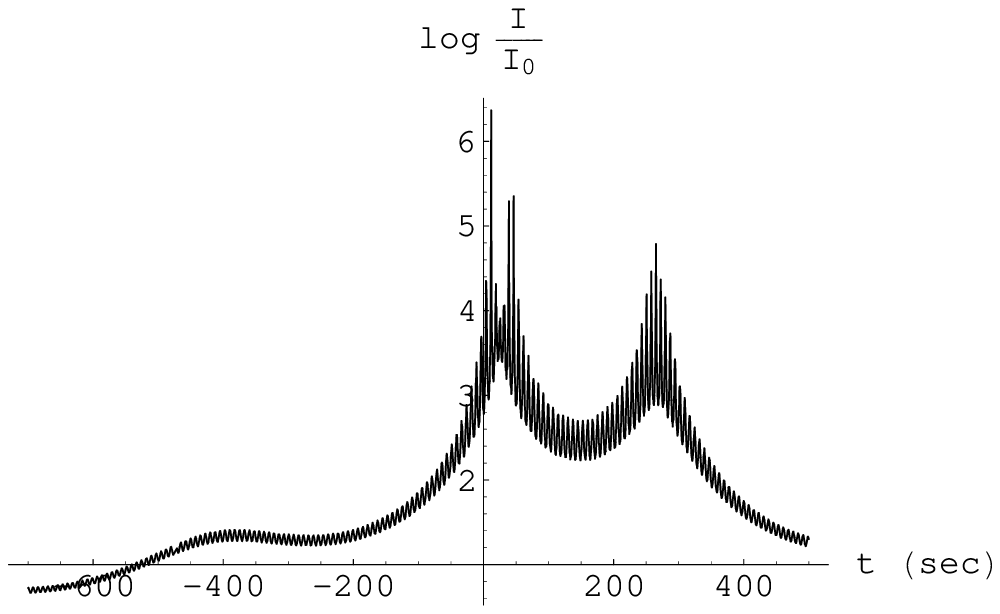}
\end{center}
\caption{A close up of the 2D trajectory of the precessing jet and
of the way it blazes the observer along the line of sight ({\em
upper panel }), and the corresponding light curve profile ({\em
lower panel }). Here we have used  the same set of  parameters
described in \S 2  and in Fig. \ref{spinning_GRB}. This set
provides a good fit to the light curve of SGR 1806-20 observed
during the giant flare (Hurley et al. 2005, Mereghetti et al.
2005). The early precursor detected $\sim$ 100 s before the onset
of the giant flare, may be due to a random trajectory of the jet
not shown in this picture.} \label{profile_SGR}
\end{figure}

\begin{figure} \begin{center}
\includegraphics[width=6.5cm,height=6.cm]{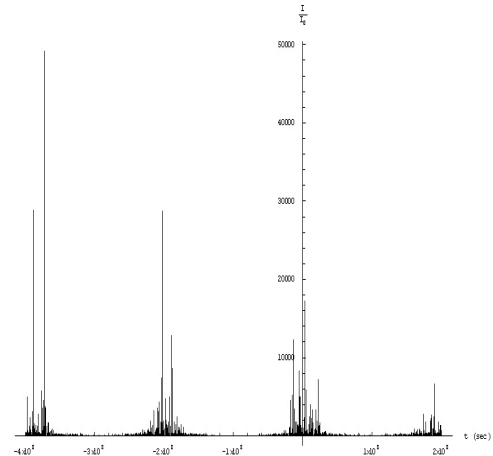}
\end{center}
\caption{A simulation of the SGR 1806-20 activity along a time
range of $\sim$ 30 yr showing a periodicity of $\sim$ 6.5 yr. This
result has been obtained with the same set of parameters discussed
in \S 5.1. The clustering of the events is related to the
precessing motion of the entire cone described by the jet, whose
period is $2 \pi/\omega_a \approx 13$ yr.} \label{SGR_timerange}
\end{figure}

Where does such an increase (7 times higher) in the energy density
come from, given that the rotational energy is at least three
orders of magnitude lower (see Eq. \ref{Erot})?

Let us go back to the intense flare activity of 2004 December 27.
No sudden change in the spin frequency ($\nu = 1/P$) was found
after the flare and an upper limit of $\Delta \nu < 2 \times
10^{-5}$ Hz has been set on the frequency variation (Woods et al.
2005). On the other hand a decrease of a factor 2.7 in the
spin-down rate has been recently observed roughly one month after
the giant flare, compared to the average value of the past 4 years
(Woods et al. 2005).
If $P \dot{P}$ is proportional to $B^2$, the decrease in $\dot{P}$
should imply from Eq. \ref{Ufin_versus_P} that the energy released
in the flare should have been $\sim 5 \times 10^{47}$ erg, while
only a tenth of this value ($\sim 5 \times 10^{46}$ erg) has been
recorded by RHESSI (Hurley et al. 2005) and GEOTAIL (Terasawa et
al. 2005). Where has the remaining  energy gone?

It follows that the evidence for a correlation between $P \dot{P}$
and $B^2$ is not at all clear thus far, especially within the
magnetar scenario.

Again, we suggest that a thin collimated jet is able to solve
these puzzles better than the magnetar model.

In Fig. \ref{profile_SGR} and \ref{SGR_timerange} we show the
results of our model (see Eq. \ref{GRB_Equation}) with the set of
parameters displayed in Table \ref{jet_parameters}. First, a thin jet would be able 
to lower the total energy budget of the process ($\dot{E}_{SGR}
\simeq \frac{\Omega}{\Delta \Omega} \dot{E}_{psr} \simeq 10^9
\times 10^{38}$ erg s$^{-1}$). Secondly, a jet in precession would
provide a natural interpretation of the features observed in the
light curve of SGR 1806-20 as one can see from
Fig. \ref{profile_SGR}.

The  structure of the $\gamma$ jet (originated either via ICS or
synchrotron radiation) consists, because of the relativistic
kinematic, in concentric cones where the inner and more collimated
ones correspond to the more energetic gamma radiation. A section
of this cone perpendicular to the axis would show different
concentric rings. If the motion of the jet is slow\footnote{Fast
and slow  jets are defined by the ratio $v_{\bot}/c$ ($>$ or $<$ 1
respectively) where $v_{\bot} = \omega_{psr} R_L \sin
\theta_{psr}$ and $R_L \sim c/P$}, the profile of the light curve
would appear symmetric to an observer, whereas if the jet is fast,
the inner structures (at higher energy) move quicker than the
outer ones (at lower energy). This determines a "compression" of
the high energy cones along the direction of the precessing
motion, the faster the motion of the jet the stronger this effect.
Fast moving jets though should appear  with a fast-rising hard
signal, followed by a lower energy tail. Consequently, a fast thin
jet that suddenly blazes the observer along its precessing motion
may explain the short e-fold time of the initial rise ($\Delta
t\sim 1.3$ ms; Terasawa et al. 2005). The  sudden oscillatory
modulation detected along the decay profile at 61 ms $<t<$ 170 ms
and 397 ms $<t<$ 500 ms (Terasawa et al. 2005; Palmer et al. 2005)
finds explanation if we assume  $\Delta t \simeq P/\gamma \simeq
1.3 \cdot 10^{-3}$ s for $P = 7.5$ s and $\gamma \sim 10^4$.

We imagine that the rotational energy losses of the pulsar are
mainly related to the braking torques of an accretion disc which
determine the change of the period derivative. This implies that
$P \dot{P}$ is not correlated to the magnetic energy density
(contrary to the magnetar model) before and after the flare.

The recent increase of the hardening might be due to the motion of
the cone described by the precessing jet: the more it bends
towards the observer the longer is its $\dot{P}$, the
more frequent may be the SGR activity and the harder is the spectum (Mereghetti et al. 2005). 

Moreover the observations of the radio afterglow of the SGR
1806-20 flare (Taylor et al. 2005) give evidence for a
polarization variability, for a rebrightening in the radio light
curve (Gelfand et al. 2005) and they show a variation in the
position of the centroid of the afterglow, that appears as
asymmetric and elongated: all these properties are in favour of
the jet interpretation. In this case a wider and less collimated
cone of GeV elecrons may produce the radio emission of the
afterglow.

In our model, highly energetic particles  are ejected
from the poles of a spinning neutron star. 
Here we propose three possible scenarios that can account for the
gamma emission from SGRs, particularly for what has been observed
in the $\gamma$-ray  giant flare from SGR 1806-20. (1) In the
simplest case the jet is made of   GeV electrons that emit
$\gamma$ rays via ICS onto thermal photons, as we have already
discussed in previous works.  Then we examine the possibility that
the soft gamma flare results from synchrotron radiation (due to
the galactic magnetic field) of PeV electrons possibly originated
by either (2) a primary jet made of muons, or (3)  by EeV nucleons
($p + \gamma \rightarrow \pi + n; \pi \rightarrow \mu \rightarrow
e$; $n \rightarrow p + e + \bar{\nu}_e$). At this stage we cannot
decide which model is able to best interpret the giant flare
emission. PeV muons have the advantage of escaping for about 100
seconds before they decay. 
On the other hand the introduction of an
hadronic component of the jet might be necessary, given  the
evidence for a spatial correlation between an excess of EeV CRs
and the location of SGR 1806-20.

\subsection{Gamma jets by IC of GeV electrons}

As we proposed in previous works, the simplest approach is to
assume that the gamma emission arises from Inverse Compton
Scattering of GeVs electron pairs ($\gamma_e \geq 2 \times 10^3$)
onto thermal photons (Fargion 1995, 1996, 1998, 1999).  Their ICS
will induce an inner jet whose angle is $\Delta \theta < 1/\gamma
\sim 5 \cdot 10^{-4}$ rad $\sim 0.0285^{\circ}$ and a wider, less
collimated X, optical cone. 
Indeed the electron pair Jet may generate a secondary beamed
synchrotron radiation component at radio energies, in analogy to
the behaviour of BL Lac blazars whose hardest TeV component is
made by ICS, while its correlated X emission is due to the
synchrotron component. Anyway the inner jet is dominated by harder
photons while the external cone contains softer X, optical and
radio waves.  In a first approximation the gamma emission is given
by the Inverse Compton relation: $< \epsilon_{\gamma} > \simeq
\gamma_e^2 kT$ for $kT \simeq 10^{-3} - 10^{-1}$ eV, and $E_e
\sim$ GeV, leading to the characteristic X spectrum. However the
main puzzle of this scenario is the way of originating such high
energy particles and how they can escape from the strong magnetic
field of the pulsar. Other mechanisms and progenitors need to be
introduced to guarantee the escape from the star of the high
energy electrons forming the jet.

\subsection{PeV Muon jets in SGRs ?}

Similarly to the GRB case, we assume a primary jet made of muons
with $E_{\mu} \sim$ 1 - 10 PeV, that can travel for about 100 s
before they decay into electrons (see Fig.
\ref{interaction_length}). The presence of a UV/optical photon
background that can prevent the propagation
of high energy electrons is negligible around a pulsar. 

We are also assuming that the jet is aligned with the magnetic
lines, thus muons would not emit synchrotron photons as they move
through the very intense magnetic field of the pulsar  generally
expected near its surface (possibly as high as  $10^{15}$ G). The
advantage of a primary beam of muons is that it guarantees that
the electrons are produced away from the surface of the neutron
star, where the magnetic field would be much lower.

When electrons of energy $\sim 1 PeV$ arise from the decay of
muons after about 100 s, the pulsar magnetic field at this
distance has decreased by a factor $B_{15} (R_{psr}/r)^{-3}$ G
$\simeq 3 \times 10^{-5} B_{15} (R_{psr}/10km)^{-3}$ G, with
$B_{15} = (B/10^{15}$ G$)$. In practice $B$ at 100 s is comparable
to the galactic magnetic field. Given that the synchrotron losses
for $B \sim 10^{-5} - 10^{-6} \, G$ acts on a fairly long time
scale, it is more likely  that PeV electrons lose most of  their
energy while they propagate through the galaxy and they interact
with the galactic magnetic field. This would lead to a radiation
of energy


\begin{equation}
E_{\gamma}^{sync} \simeq 1.7  \times 10^5 \left(\frac{E_e}{10^{15}
\: eV} \right)^2 \left(\frac{B}{2.5 \cdot 10^{-6} \; G} \right) \:
eV \label{sync_SGR_muon}
\end{equation}


with  an energy loss distance scale of



\begin{equation}
 t^{sync}    \simeq  6.3 \times 10^{10}
  \left(\frac{E_{e}}{10^{15} \, eV} \right)^{-1}
\left(\frac{B}{2.5 \cdot 10^{-6} \, G} \right)^{-2} \: s
\label{lsync_SGR_muon}
\end{equation}


The Larmor radius for the galactic magnetic field is

\begin{equation}
\frac{R_L}{c} \simeq 8.2 \times 10^{7}
  \left(\frac{E_{e}}{10^{15} \, eV} \right)
\left(\frac{B}{2.5 \cdot 10^{-6} \, G} \right)^{-1} \: s
\end{equation}

and it is about three orders of magnitude lower than the
synchrotron energy loss distance ($R_L/ct_{sync} = 1.3 \cdot
10^{-3}$). This may cause a widening of the  opening angle of the
jet, that would have a final "fan"-like  (see also Fargion 2002,
Fargion et al. 2004) or a double disc where the $e^{\pm}$ are
bounded by the magnetic field. The jet would open along a two
dimensional plane, given that the galactic magnetic field is
homogeneous on a scale of 55 pc (Protheroe 1990). We are assuming
a Lorentz factor which is $\gamma_e \sim 2 \times 10^{9}$, but in
this case the solid angle of the jet is not $\propto 1/ \gamma^2
\sim 2.5 \times 10^{-19}$, but $1/ \gamma \times \theta_L$, where
$\theta_L$ is the aperture angle due to the Larmor precession
around the galactic field. For $\theta_L \simeq 10^{\circ}$,
$\Delta \Omega \sim  10^{-10}$, and this aperture angle guarantees
that the number of events is not suppressed by a too narrow
beaming.


\subsection{ EeV nucleons as  progenitors of  SGR $\gamma$ jet?}

\begin{figure}
\begin{center}
\includegraphics[width=8cm,height=6cm]{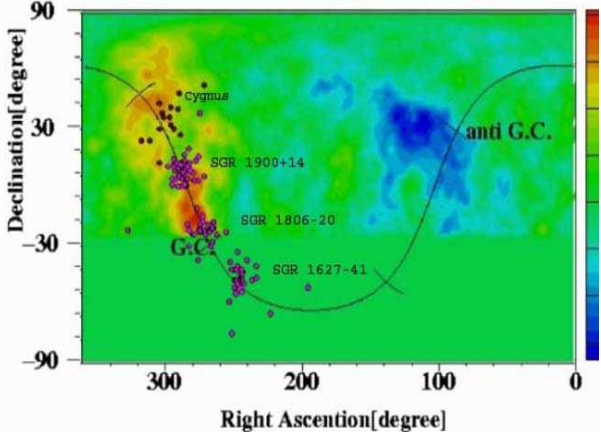}
\end{center}
\caption{The correlation between the BATSE data, the AGASA
discovery of an anisotropy in the arrival direction of EeV CRs
near the galactic center and the  Cygnus regions and the position
of the galactic SGRs. The four clusters of the data from BATSE in
the left-hand side of the map correspond from top to bottom to:
the Cygnus region, SGR 1900+14, SGR 1806-20, SGR 1627-41. Note
that AGASA was unable to record of the last SGR because below its
horizons.The fifth SGR 0526-66 in the Magellanic cloud is not
indicated in this map because it did not show signs of activity
between 1997 and 2000. Most of the authors Hayashida et al. (1999)
or Fargion (2002) argued that this excess in the CR distribution
might be due to EeV neutrons partially decaying in flight.}
\label{EEV-SGR}
\end{figure}

Finally we consider  the possibility that SGRs may be sources of
UHECRs, and that
the PeV muons may be originated by jets made of higher energy
hadrons. Similar ideas on UHECRs  have been recently suggested
also by other authors (Asano, Yamazaki, \& Sugiyama 2005), but
here we assume that the energy of the hadrons from SGRs does not
exceed $10^{18}$ eV instead of being around $10^{20}$ eV as
proposed by Asano and collaborators. Indeed there is strong
evidence in CR spectra that the origin of EeV particles may be
both galactic and extragalactic. Excluding very rare SN/GRBs, in
our galaxy there are no active sources
 at this range of energies, thus we cannot ignore the role of SGRs as possibly related to EeV CRs. 

In 1999 AGASA reported the measure of a significant anisotropy
(4\%) in the arrival direction of CRs in the small energy range
$10^{17.9} - 10^{18.3}$ eV as due to the excess of events in two
regions of about 20$^{\circ}$ near the SGR 1806-20 and Cygnus, and
it has been argued that this anisotropy was due to EeV neutrons
(Hayashida et al. 1999). Interestingly enough we have also found a
correlation between the BATSE data of SGR 1806-20 and the AGASA
excess, as one can see from Fig. \ref{EEV-SGR}, where we have
overlaid the BATSE detections (between 1997 and 2000)
to the AGASA map. This
would suggest that the neutron signal and the soft-gamma detection
from BATSE are related to the same mechanism.

To explain such an interesting correlation, we will start with the
assumption that protons can be accelerated by the large magnetic
field of the pulsar up to EeV energy. The protons could emit
directly soft gamma rays via synchrotron radiation with the
galactic magnetic field ($E_{\gamma}^p \simeq 10 (E_p/EeV)^2
(B/2.5 \cdot 10^{-6} \, G)$ keV), but the efficiency is poor
because of the too long timescale of proton synchrotron
interactions. Photopion production must occur to produce the
observed neutron excess, a process that also gives birth to
neutral and charged pions. Which photons represent the target for
such a process is not completely clear yet. It is possible that
the object may have an IR emission and that this thermal
background acts as a target for the high energy protons. In the
literature there are claims of identification of an IR counterpart
to SGR 1806-20. Eikenberry et al.  (2001) report the discovery of
possible IR counterparts to the object, and more recently Israel
et al. (2005) claim the detection of one faint object ($K_s = 21.6
\pm 1.0$) which is compatible with the radio position of the SGR
(Gaensler et al. 2005) obtained with a very accurate positional
accuracy (of
0.1\arcsec). 
The energy of the thermal photons necessary to produce pions is
$\epsilon_{\gamma} = 0.2$ GeV$^2/E_p$ = 0.2 eV for $E_p = $
10$^{18}$ eV. Charged pions (born with roughly a tenth of the
original energy of the proton) decay into muons and then into
electrons with $E_e \lesssim 10^{16}$ eV. Medina-Tanco \& Watson
(2001), following Stecker (1968) argues that during the
propagation of EeV protons from the galactic centre,  photopion
production with ambient galactic IR photons ($p + \gamma_{IR}
\rightarrow \Delta^+ \rightarrow n + \pi^+$) is favoured compared
to proton proton collisions ($p + p \rightarrow n + p + N\pi$) in
order to produce charged pions, and they obtain a mean free path
of 37 pc for the interaction with an IR galactic photon background
($T_{IR} \sim 100$ K). Thus, even if there is not an IR source
localised with the SGR, the interaction with the galactic IR
background may be efficient in converting the protons into pions
and neutrons (see also Grasso \& Maccione 2005).

Pions then decay into muons that decay into high energy electrons
($E_e \lesssim 10$ PeV) that, interacting with the local galactic
magnetic field, lose energy via synchrotron radiation:

\begin{equation}
E_{\gamma}^{sync} \simeq 4.2  \times 10^6 \left(\frac{E_e}{5 \cdot
10^{15} \: eV} \right)^2 \left(\frac{B}{2.5 \cdot 10^{-6} \; G}
\right) \: eV
\end{equation}

with an interaction length given by



\begin{equation}
 t^{sync}    \simeq  1.3 \times 10^{10}
  \left(\frac{E_{e}}{5 \cdot 10^{15} eV} \right)^{-1}
\left(\frac{B}{2.5 \cdot 10^{-6} \, G} \right)^{-2} \: s
\end{equation}

The Larmor radius is about two orders of magnitude smaller than
the synchrotron interaction length and this may imply that the
aperture of the jet is spread by the magnetic field.

\begin{equation}
\frac{R_L}{c} \simeq 4.1 \times 10^{8}
  \left(\frac{E_{e}}{5 \cdot 10^{15} eV} \right)
\left(\frac{B}{2.5 \cdot 10^{-6} \, G} \right)^{-1} \: s
\end{equation}

In this "hadronic" scenario we are assuming  a Lorentz factor
which is $\gamma_e \sim \times 10^{10}$, and, as mentioned in the
previous section, the solid angle of the jet is not $\propto 1/
\gamma^2 \sim \times 10^{-20}$, but $1/ \gamma \times \theta_L$,
where $\theta_L$ is the aperture angle due to the Larmor
precession around the galactic field. For $\theta_L \simeq
10^{\circ}$, $\Delta \Omega \sim  10^{-11}$, and this aperture
angle again guarantees that the number of events is not suppressed
by a too narrow beaming.



The photopion interactions produce also neutrons at $\sim 10^{18}$
eV ($\gamma \sim 10^9$), which have a decay length of $\lesssim$
10 kpc (E/$10^{18}$ eV), and they can easily propagate along these
distances without being bent by the galactic magnetic field.
Neutron beta decay produces a proton, electron and an electronic
antineutrino. Given its long decay length, the production of the
electron would occur well far away from the original pulsar. The
electron is produced with an energy spectrum that ranges between
$10^{15}$ eV $< E_e < 10^{17.5}$ eV and peaks at $E_e \sim
10^{15}$ eV. Again PeV electrons interact with the local galactic
magnetic field producing  via synchrotron radiation photons of few
hundreds keV as given by Eq. \ref{sync_SGR_muon}.


For higher electron energies, around $E_e \gtrsim 10^{16.5}$ eV
the Larmor radius becomes comparable and even larger than the
synchrotron interaction length, 
therefore the opening angle of the jet would remain collimated.
However such an angle would be too small and would make SGR events
extremely unlikely. Again for the higher energy electrons the peak
energy of synchrotron photons would be

\begin{equation}
E_{\gamma}^{sync} \simeq 1.7  \times 10^9 \left(\frac{E_e}{10^{17}
\: eV} \right)^2 \left(\frac{B}{2.5 \cdot 10^{-6} \; G} \right) \:
eV
\end{equation}

and to explain the hundreds keV emission we have to assume that
the jet is not pointing exactly towards the Earth, but it is
slightly off-axis so that only the softer range of frequencies
could have been observed during the flare.

\section{Conclusions}

The most popular models to date  to interpret GRBs and SGRs, the
fireball collimated in a few degrees jet for GRBs, and the
spherically symmetric magnetars for SGR, cannot completely explain
all the issues that these events do present.

We have proposed an alternative scenario  by introducing a model
with highly collimated jets of high energy particles in
precession. We are assuming the energy equipartition between the
power of the jet and that of the SN (for GRB) or the X-ray pulsar
(for SGRs). The gamma rays emerge in a small opening angle, and
the narrow beaming allows to dramatically reduce the total output
power (from $ \simeq 10^{50}$ erg s$^{-1}$ for the fireball/jet
model of GRBs to 10$^{43}$ -- 10$^{44}$ erg s$^{-1}$ in our GRB
case as the typical SN power, and from $10^{47}$ erg s$^{-1}$ for
the magnetar to $10^{38}$ erg s$^{-1}$ in our SGR scenario,
comparable the the observed power of  X-ray pulsars), which is the
most radical assumption in such isotropic models.

Compared to our previous versions  here we have considered that
high energy electrons are the decay product of a primary particle
that in the case of the GRB is a muon.  For SGRs we have
considered  a jet of either prompt muons or of protons that
produce, through photopion production, secondary muons and
electrons. This proposal arises from the fact that the propagation
of a  jet of prompt electron pairs would be severely suppressed by
the extreme opacity conditions  in the proximity of the surface of
a pre-SN star or a highly magnetised pulsar.

If GRBs are originated in a core collapse of very massive stars, a
jet of muons would have a higher chance to escape from the stellar
surface. Secondary electrons produced after the decay of the muons
at about 100 s from the surface, may produce the gamma emission
after a long chain of interactions with the stellar background
radiation (ICS), which gradually reduce their initial energy, from
$\sim 10^{15}$ eV to a tens of GeV. A narrow jet would blaze the
observer at high energy when viewed on-axis (at viewing angle
$\sim 1/\gamma$), while it would appear "softer" if observed
off-axis. This could give an explanation to the wide range of
properties and emission from different GRB events.  A larger
viewing angle would be associated to the softer emission of the
X-ray afterglow, and it may also be linked to the properties of
objects such as the XRFs.

A similar interpretation has been introduced for SGRs, with a
particular attention to the giant flare from SGR 1806-20 on 2004
December 27. We have discussed different  scenarios at the origin
of this event. We have shown how a primary jet of muons (with
$E_{\mu} \simeq 10^{15} - 10^{16}$ eV) decaying into high energy
electrons, may be a source of a collimated gamma radiation. In
this case the favoured mechanism of the emission would be the
synchrotron interactions with the galactic magnetic field, since
the secondary electrons would be produced at a distance from the
pulsar where its magnetic field would be negligible. 
The interactions with the galactic magnetic field will widen the
shape of the jet like a "fan" (see also Fargion  2002, Fargion et
al. 2004), and its solid angle would still be very narrow $\Delta
\Omega \propto 1/\gamma \sim 10^{-10} - 10^{-11}$. In this
scenario the PeV leptons ($\mu^{\pm}$ decaying into electron
pairs) might also  be the secondaries of EeV protons that generate
charged pions and neutrons via photopion production with the
galactic IR background (Medina-Tanco \& Watson 2001; Grasso \&
Maccione 2005). We have considered that  SGRs may be sources of
EeV protons and neutrons, on the basis of the AGASA discovery of
an anisotropy in the UHECR signal in this energy range (Hayashida
et al. 1999). This excess seems to be related to the area where
SGR 1806-20
 is located (Bellido et al. 2001) and  we also found a correlation
between, the AGASA excess, the position of SGR 1806-20, SGR
1900+14, Cygnus and the BATSE public data that recorded the
activity of these objects between 1997 and 2000 (see Fig.
\ref{EEV-SGR}).



We foresee that if the role of nucleons  as primaries of the soft
gamma emission of SGR 1806-20 is correct, the giant flare might
also be source of a prompt (and possibly repeating)
$\gamma$-induced shower (made by $\gamma$ photons from the decay
of PeV neutral pions) that may have been detected by Milagro in
correspondence with the 2004 December 27 flare. We also expect
that a rich component of the EeV neutrons (or protons from their
decay) might appear 
in the AUGER or HIRES detectors, in rough coincidence with this
event. Because of the delayed  arrival time of protons, one should
expect also a long and persistent UHECR afterglow. Finally a
signal of secondary muons at PeV energies, induced by high energy
neutrinos from the SGR, might occur in Amanda.

To conclude, we imagine  that if the precessing jet model gives a
correct interpretation  of the properties of SGRs, SGR 1806-20
will  still be active in the next months and years.

\end{document}